Corresponding Author: Professor Jianbo Wang, Ph.D

Corresponding Author's Institution:

First Author: Zhi Ma, Ph.D

Order of Authors: Zhi Ma, Ph.D; Jing Yuan; Chentao Cao; Qingfang Liu, Ph.D ; Jianbo Wang, Ph.D

Abstract: Ni coated SiO2 and SiO2 coated Ni composite particles with core-shell structures were designed, prepared and their microwave absorption properties were characterized. The comparison study of the shell effect on the effective electromagnetic parameters reveals that the effective permittivity/permeability was crucially determined by the percent and the intrinsic electromagnetic parameters of the component materials regardless of various core/shell structures. Both of the composite core-shell structures could have good microwave absorption properties. Investigation of the input impedance indicates that, good microwave absorption performances are a consequence of proper electromagnetic impedance matches when the effective electromagnetic parameters were modulated.

Suggested Reviewers: zhaohua Cheng
Institute of Physics, Chinese Academy of Sciences
zhcheng@aphy.iphy.ac.cn
He is an expert in this area.

zhengwen li
Temasek Laboratories, National University of Singapore
tsllizw@nus.edu.sg
He is an expert in the microwave absorption field.

zhidong zhang
Institute of Metal Research, Chinese Academy of Science
zdzhang@imr.ac.cn
prof. zhang zhi-dong is famous in this field.We knew that he is very severe and fastidious to manuscripts, we hope he will give us more suggestions to improve ours English and quality of the manuscript.

Opposed Reviewers:

**Cover Letter**

*Materials Chemistry and Physics (letter)*

Dear Editor,

We send our manuscript "**Comparative study of microwave absorption in Ni/SiO$_2$ and SiO$_2$/Ni core-shell structures**" to you and the authors list as following: **Zhi Ma, Jing Yuan, Chentao Cao, Qingfang Liu and Jianbo Wang**. The corresponding author is **Jianbo Wang and Qingfang Liu**.

We promise the manuscript has not been previously published, is not currently submitted for review to any other journal, and will not be submitted elsewhere before a decision is made by your journal.

We earnestly hope that it could be accepted for publication in your journal of *Materials Chemistry and Physics*. Please notify us if there is any further question. Thank you very much for your time in reviewing this manuscript. We look forward to hearing from you very soon. Any of your early response will be highly appreciated. Thank you very much.

Sincerely yours,

Jianbo Wang and Qingfang Liu

June 23, 2011

Professor Jianbo Wang and Qingfang Liu
Key Lab for Magnetism and Magnetic Materials of the Ministry of Education,
Lanzhou University,
Lanzhou 730000, People's Republic of China
Email: **wangjb@lzu.edu.cn (J. B. Wang), liuqf@lzu.edu.cn (Q. F. Liu)**
Tel: +86-931-8914171
Fax: +86-931-8914160

*Materials Chemistry and Physics*

Title: Comparative study of microwave absorption in Ni/SiO$_2$ and SiO$_2$/Ni core-shell structures

Corresponding Author: Professor Jianbo Wang

Dear Editor,

Thank you for your useful comments and suggestions on the language and structure of our manuscript. We have modified the manuscript accordingly, and detailed corrections are listed below point by point:

1) Each point provided in 'Research highlights' should be represented by bullets and should not exceed a maximum of 85 characters including spaces.

    We have represented the WHOLE 'Research highlights' carefully according to the Guide for Authors and make sure that each bullet point not exceed a maximum of 85 characters including spaces or maximum 20 words. The 'Research highlights' were revised and extended to 4 points and the words counts are now 13, 19, 13 and 14, respectively. We believe that the 'Research highlights' is now acceptable for the review process.

2) Please note that the reference list must conform strictly to the Guide for Authors. For journal articles, only starting page number should be provided.

    We have checked all the references and formatted them strictly according to the Guide for Authors. Especially, ONLY journal starting page number has now been provided.

The manuscript and the 'Research highlights' file have been resubmitted to your journal. We look forward to your positive response.

Sincerely,

Jianbo Wang

June 29, 2011

## Highlights:

1. Ni/SiO$_2$ and SiO$_2$/Ni composite particles with core-shell structures were designed and characterized.

2. The comparison reveals that effective permittivity/permeability was crucially determined by percent of component materials regardless of various structures.

3. Good microwave absorption performances are a consequence of proper electromagnetic impedance matches.

4. Proper electromagnetic impedance matches could be obtained when effective electromagnetic parameters were modulated.

# Comparative study of microwave absorption in Ni/SiO$_2$ and SiO$_2$/Ni core-shell structures

Zhi Ma, Jing Yuan, Chentao Cao, Qingfang Liu\*, Jianbo Wang\*

*Key Laboratory for Magnetism and Magnetic Materials of Ministry of Education, Lanzhou University,*

*Lanzhou 730000, People's Republic of China*

**Abstract**

Ni coated SiO$_2$ and SiO$_2$ coated Ni composite particles with core-shell structures were designed, prepared and their microwave absorption properties were characterized. The comparison study of the shell effect on the effective electromagnetic parameters reveals that the effective permittivity/permeability was crucially determined by the percent and the intrinsic electromagnetic parameters of the component materials regardless of various core/shell structures. Both of the composite core-shell structures could have good microwave absorption properties. Investigation of the input impedance indicates that, good microwave absorption performances are a consequence of proper electromagnetic impedance matches when the effective electromagnetic parameters were modulated.

\* Corresponding authors.

Tel.: +86 931 8914171; fax: +86 931 8914160.

*E-mail address:* wangjb@lzu.edu.cn (J. B. Wang), liuqf@lzu.edu.cn (Q. F. Liu).

## 1. Introduction

Nowadays, a great deal of interest has been attached to the development of magnetic composites for their unique nanostructure and microwave absorptive properties [1-3]. Moreover, the low-cost, low-density dielectric-magnetic composites are expected to be used in commercial and military industries, since the heavy conventional absorptive materials have difficulties in applications that requiring lightweight mass and high reflection loss in GHz frequency [4]. Meanwhile, the asymmetric permittivity or permeability of absorbers damage impedance match and thus inhibit their practical applications in microwave absorption. Therefore, it is pressing to explore the solutions of the above problem and promote the application of core-shell structured dielectric-magnetic composites.

Recently, the core-shell structured composites and their microwave applications have been studied extensively [5]. Among the candidates for electromagnetic (EM) wave absorbers, magnetic composites, such as magnetic particles coated with a nonmagnetic insulator or dielectric particles coated with a magnetic shell, have attracted particular interest [6]. Zhang and co-workers reported that carbon-coated FeNiMo nanocapsules exhibited good microwave absorption properties, showing that an optimal reflection loss (RL) of −64.0 dB was obtained at 13.2 GHz with an absorber thickness of 1.9 mm [7]. For carbon-coated Fe nanocapsules, the absorption range for exceeding -10 dB is in 2–18 GHz for the absorber thicknesses of 1.54–7.50 mm [8]. Zhen and co-workers reported that a RL about −75 dB as well as a super broad absorption band (exceeding -20 dB about 11 GHz) can be simultaneously obtained for Fe/SiO$_2$-based coatings [9]. Additionally, compared to magnetic/dielectric core/shell structured composites, the magnetic materials coated nonmagnetic insulator core-shell materials also exhibited good microwave absorption properties. In our previous work [10], it was found that Ni–Co–P coated SiO$_2$ powder possesses excellent microwave absorption properties. The maximum microwave loss reaches -48 dB at 4.2 GHz with a thickness of 3.10 mm. In this work, the maximum microwave loss of Ni coated silica powders is less than -40 dB at the frequency of 9.3 GHz with a thickness of 2.90 mm. The RL values are less than -20 dB in the range of 7.9-12.3 GHz over absorber thicknesses of 2.3-3.4 mm. These phenomenons are inconsistent with the conventional view that metallic magnetic shell structures which have a high permittivity will damage the impedance matches and thus a poor microwave absorption performance is destined.

However, for the lightweight microwave absorbers, it is difficult to find the optimum conditions because there are six EM parameters (i.e., $\mu'$, $\mu''$, $\varepsilon'$, $\varepsilon''$, $d$, $f$) which should be considered. On the other hand, the input EM impedances of the dielectric-magnetic core-shell composites were crucially modulated by the component materials and their diameter/thickness. The comparative study on the microwave absorbing and EM impedance match of core/shell composite coatings is still not reported. All of these encourage us to study the EM impedance matches and microwave absorption in the opposite core/shell structures. In this work, two opposite core-shell structures were designed and studied. Good microwave absorption properties of both the different structures were demonstrated.

**2. Experimental**

2.1. Preparation of Ni coated silica and $SiO_2$ coated Ni core-shell powders

Monodisperse silica particles were prepared by the modified stöber method [10]. Then, Ni coated silica powders were obtained by the electroless plating method. Before plating, pretreatments including surface coarsing treatment and activating treatment were performed to improve the surface properties of silica microspheres. The coarsing treatment was performed by etching silica microspheres in 10% HF solution under the ultrasonic stirring for 30 s, and then the solution was rinsed with distilled water twice and dried. Activating treatment was carried out in a $AgNO_3$ solution [11] at 30-40 $^o$C for 60 min, the activated silica microspheres were separated from the activating solution, rinsed with distilled water twice and immersed in the sodium hypophosphite solution for 5 min, then rinsed with distilled water and dried. After these operations, the particles became brown suggesting the presence of Ag on their surface.

Electroless nickel was plated using a sodium hypophosphite bath. All the chemical reactants were analytical grade and used as received. Nickel sulphate is the source of metal ions, sodium hypophosphite is used as the reducing agent, sodium citrate is used as a complexing agent, and ammonium sulphate acts as a buffering agent to control *p*H of the bath during the plating process. The proportion of pre-treated $SiO_2$ particles that added to the plating bath was 10 g/L. The electroless plating was performed at a temperature of 90 $^o$C and a *p*H value of 8.7 by ammonia. After electroless plating, the suspension was filtered, washed thoroughly with distilled water for several times and dried. The magnetic nickel-coated silica microspheres were separated by a magnet.

Typical synthesis of Ni microspheres was carried out via a simple hydrothermal route. The major steps are similar to the preparation of Cobalt hierarchical nanostructures [12]. $SiO_2$ coated Ni composite nanostructures are prepared by a similar stöber process as it was described in [10]. Fig. 1 illustrates the Ni coated $SiO_2$ and $SiO_2$ coated Ni composite with core-shell structures.

2.2. Measurements

The surface morphologies of the prepared and pre-treated $SiO_2$ particles were investigated using a transmission electron microscopy (TEM, H-600, Japan). The scanning electron microscopy (SEM, S-4800, Japan) with EDX analysis were performed to identify the morphologies and components of the coated core-shell structures. Crystalline structure analyses of the coating were examined using an X-ray diffraction analyzer (XRD, Cu-1.54056 Å, PANalytical X'Pert, Holland). The magnetic properties of sample were studied by vibrating sample magnetometer (VSM, Lakeshore 7304). The real and imaginary part of complex permeability and complex permittivity were extracted from the measured scattering parameters, which were recorded on a vector network analyzer (Agilent E8363B) in the frequency range of 0.1-18 GHz. The measured samples contain 75 wt.% magnetic particles in the powder-paraffin composites and were made into a toroidal shape with an outer diameter of 7.00 mm and an inner diameter of 3.04 mm.

**3. Results and discussion**

3.1. Morphology and crystalline structure of the core-shell powders

Fig. 2 shows representative TEM micrographs of prepared silica spheres and pre-treated silica powders. It can be seen from Fig. 2(a) that the prepared monodisperse silica particles have a regular spherical shape with a diameter of 400 nm. Fig. 2(b) shows the silica microspheres after activating treatment. Silver nanoparticles with a diameter of 20 nm were distributed on the surface of the silica spheres. Once the surface of the silica particles is homogeneously modified, the addition of a silver salt leads to the selective redox reaction on the surface. The Ag-activation procedure assures elimination of the passive condition of the silica surfaces, which made them suitable for electroless plating.

SEM images of Ni coated $SiO_2$ particles (sample A) are shown in Fig. 3(a) and (b). It shows that the spherical characteristic of the silica particles is lost and the particles display irregular shape with the sizes in the range of 0.6-1.3 μm. The coating on the silica microspheres is smooth and unique except a few nubbles or holes. It is well known that the electroless plating occurs at the active centers on the silica substrates and the nickel deposition continues. In general, the onset of deposition reaction and uniformity of deposited nickel films on the silica sphere surfaces, depend on the absorption amounts of the Ag catalyst. So an increase of the active sites on the surface of silica spheres will lead to an ideal deposition of metal on the spheres. Fig. 3(c) and (d) show the surface morphologies of the prepared Ni particles and $SiO_2$ coated Ni core-shell structured particles. As shown in Fig. 3(c), the final product is composed of large and rough particles with the size about 4.5 μm. After it was coated, the image of sample B in Fig. 3(d) reveals that the surface of the spherical particles is smooth. The existence of a $SiO_2$ shell was also confirmed by the EDS spectra.

The XRD patterns of the Ni coated $SiO_2$ core-shell structures are shown in Fig. 4(a). It shows that Ag is well deposited on the surface of the silica powder after the activation procedure. Two peaks at 38.1° and 64.4° represent the well-defined peaks corresponding to the diffraction from (111), (220) planes of Ag. Three peaks at 44.5°, 51.8° and 76.3° are corresponding to the diffractions from (111), (200) and (220) planes of Ni, respectively. In Fig. 4(a), the XRD patterns consist of sharp peaks superimposed on broad peaks near 44°. The samples exhibit the characteristic of mixture of amorphous and crystalline phase [13]. Furthermore, The X-ray fits were analyzed by deconvolution the original data and fitting them with amorphous and crystalline components. The area under these fitting peaks was calculated to approximately determine the percentage of phases that constitute the sample [14]. The size of crystalline of the samples can be calculated approximately by the Scherrer equation [15]. The calculated results reveal that the size of the crystalline phases for the sample is about 4.8 nm. In Fig. 4(b), three peaks are observed in $SiO_2$ coated Ni structures, which are also corresponding to the diffractions from (111), (200) and (220) planes of Ni. No $SiO_2$ peaks were observed in this core-shell structure, which exhibits the characteristic of amorphous phase.

3.2. Magnetic properties of the core-shell structured composite materials

The hysteresis loop obtained for the core-shell structured composite materials at room temperature are shown in Fig. 5. The Ni coated $SiO_2$ particles show a saturation magnetization ($M_S$) of 26 emu/g. The $SiO_2$ coated Ni particles show a $M_S$ of 45 emu/g. However, the $M_S$ is lower than that of pure nickel metals (52 emu/g for Ni particles), which is mainly due to the presence of non magnetic materials, nanocrystalline grains and grain boundaries [16, 17]. The $H_C$ of Ni coated $SiO_2$ particles is determined as 7 Oe, which is close to 13 Oe for Ni particles, and much lower than 101 Oe for $SiO_2$ coated Ni particles. $H_C$ of the $SiO_2$ coated Ni particles is 14 times as large as that of Ni coated $SiO_2$ particles. The enhancement of the coercivity is due to the effect of surface spin pinning in hierarchical structures [18]. Owing to the fact that $M_S$ should be proportional to the magnetic weight percent, the shell thickness of the composite could be calculated. For $SiO_2$/Ni core/shell structures, the $M_S$ of sample A decreased to 50% as low as that in pure Ni particles and the core diameter is about 400 nm, thus the thickness of Ni shell about 35 nm could be estimated. Also, due to the $M_S$ percent of sample B is 86.5% and the core diameter is 4.5 μm, the calculated shell thickness for Ni/$SiO_2$ core/shell structures is about 700 nm. Compared to the core diameters, the shell thicknesses of the composite particles are very thin.

3.3. Effective permittivity and permeability of the core-shell composite particles

The complex permittivity/permeability is shown in Fig. 6. As illustrated in Fig. 6(a), Ni coated $SiO_2$ particles exhibit that the real part of the complex permittivity ($\varepsilon'$) decreases with increasing frequency except two resonance peaks in the region of 8.0-15.0 GHz. However, the imaginary part ($\varepsilon''$) of complex permittivity increases with increasing frequency and exhibits two distinct resonance peaks around 11.0 and 14.0 GHz. The complex permittivity of $SiO_2$ coated Ni particles are also shown in Fig. 6(a). The permittivity spectra show the same behaviors but lower values within the measured frequency range. In general, the dielectric resonance provides that there exists space charges in the materials. High-frequency resonance is attributed to atomic and electronic polarization [19]. This dielectric resonance behavior is expected when the composite is conductive and the skin effect becomes significant. In our studies, the resonance frequency of the powders is related to the high conductivity of nickel.

Actually, a strong effect of the shell on the effective parameters, just as the effective permittivity and permeability, and thus on the electromagnetic impedance matches, is not a priori apparent. Also it is not clear how the wavelength of microwaves affects the effective permittivity/permeability. So it is difficult to compare the effective permittivity/permeability in two samples when they have the different dielectric and magnetic properties with these opposite structures. However, if the effective permittivity/permeability could be successfully presented as a function of the percent of component and their EM constant in the core-shell structures, this comparison will become an effective way to discuss the shell effect on the effective EM parameters. According to the sketch in Fig. 1, the effective dielectric permittivity of the core-shell structures [20] could be described approximately by:

$$\varepsilon_{eff} = \frac{\varepsilon_1 \varepsilon_2}{[1-(r_1/r_2)^3]\varepsilon_1 + (r_1/r_2)^3 \varepsilon_2} \qquad (1)$$

Where $r_1$ and $r_2$ are the radiuses of core and the composites, respectively. When considering the composition and density for the core and shell materials, the above equation could be rewritten as

$$\varepsilon_{eff} = \frac{\rho_1(1-x) + \rho_2 x}{(\rho_1/\varepsilon_2)(1-x) + (\rho_2/\varepsilon_1)x} \qquad (2)$$

Where $x$ is the core weight percent, $\rho_1$ and $\rho_2$ are the densities of the core and shell materials, respectively. If $x$ were set as 0.5 and 0.865 for sample A and B, respectively, we found that the effective permittivity of sample A is larger than that of sample B. The theoretic conclusion was in agreement with the practical dielectric measuring results.

For the microwave absorbers, the imaginary part of the complex permeability ($\mu_r = \mu' - j\mu''$) plays the most important role because the microwave energy is mainly attenuated by means of the magnetic loss. It is reasonable that both the dielectric loss and the magnetic loss can be influenced by the "core-shell" microstructure of microwave absorbent. The silica powders are encapsulated in a Ni shell to form the core-shell structure. The introduction of silica powder can improve the dielectric loss of composites, and will lead to the increase of the number of dangling bond atoms and unsaturated coordination. These variations lead to the interface polarization and multiple scatter, which is helpful to absorb more microwaves. The core of silica particles enhances the dissipation of electromagnetic energy while the magnetic layer causes the demagnetizing field [21]. When Ni powders are coated with a silica shell, the $SiO_2$ shell may cause the reduction of complex permittivity and thus proper impedance match can be reached. Compared with Ni coated $SiO_2$ powders (see Fig. 6(b)), the permittivity of $SiO_2$ coated Ni powers decreases dramatically, while permeability increases slightly.

Indeed, an incident electromagnetic wave induces microscopic currents in grains which are known to cause magnetization and result in the permeability of the powder [22], the effective permeability could be written approximately by

$$\mu_{eff} = (\frac{r_1}{r_2})^3 \times \mu_1 + \frac{2[1-(r_1/r_2)^3]}{2+(r_1/r_2)^3} \times \mu_2 \qquad (3)$$

Also, when considering the compositions and the densities, the above equation could be rewritten as

$$\mu_{eff} = \frac{\rho_2 x}{\rho_1(1-x)+\rho_2 x} \times \mu_1 + \frac{2\rho_1(1-x)}{2\rho_1(1-x)+3\rho_2 x} \times \mu_2 \qquad (4)$$

If $x$ were set as 0.5 and 0.865, we found that the effective permeability of sample A is lower than that of sample B. This result is consistent with the measured values. We also plot the functions from equations (2) and (4) in Fig. 7, the circle labels the effective permittivity/permeability for samples A and B. The comparison of the opposite core-shell structures concludes that the effective permittivity/permeability is mainly determined by the components percent in the dielectric/magnetic composites regardless of different core-shell structures. Since the dielectric constant of the components is in the same order and the permeability of the components is in a different order, the effective permittivity and permeability was ultimately determined by the percent and the intrinsic EM parameters of the components in the core-shell structures.

3.4. Electromagnetic impedance matches of the core-shell composite particles

An important parameter related to the reflection loss is electromagnetic impedance matches. The condition that an electromagnetic wave is completely absorbed could be described as [23]:

$$\sqrt{\mu/\varepsilon} \bullet th(\gamma d) = 1 \tag{5}$$

Where $\gamma$ is a propagation factor, $d$ is the thickness of the ideal homogeneous medium. The Smith chart of reduced input impedance $Z_{in}=\alpha+j\beta$ was used to characterize the perfect and finite matching, which is shown in Fig. 8. In fact, when the equation (5) is satisfied, the minimum reflectivity $RL_{min}$ (dB) is theoretically negative infinity, which is known as the perfect matching. Meanwhile, at the perfect matching thickness, $Z_{in}$ passes the point at $\alpha=1$ and $\beta=0$ in the smith chart. The finite matching are calculated when the input impedance $Z_{in}$ was expressed as [24]

$$Z_{in} = \sqrt{\frac{\mu_r}{\varepsilon_r}} \tanh[\frac{\pi f d}{c} \sqrt{\mu'\varepsilon'}(\frac{\mu''}{\mu'}+\frac{\varepsilon''}{\varepsilon'}) + j\frac{2\pi f d}{c}\sqrt{\mu'\varepsilon'}]$$

(6)

Where $\mu_r = \mu' - j\mu''$, $\varepsilon_r = \varepsilon' - j\varepsilon''$. $f$ and $d$ are the matching frequency and thickness of the materials, respectively. $c$ is the velocity of electromagnetic waves in free space. The matching characteristics are significant for composite with different matching thicknesses. In the smith chart (Fig. 8(a)), as thickness $d$ increases from 2.3 to 4.0 mm, the input impedance $Z_{in}$ increases from $\alpha < 1$ to $\alpha > 1$. At the neighborhood of thickness of 2.9 mm, as the thickness $d$ increases or decreases, the resistance $\alpha$ deviates even more from the point of $\alpha=1$ and $\beta=0$. This means that, at this matching thickness (2.9 mm), the reflection loss is the minimum but a finite value. For silica coated Ni particles, as shown in Fig. 8(b), we can see that the matching thickness is 2.47 mm, at which the reflection loss reaches its minimum. The comparison of the input impedance for two samples reveals that sample A has much smaller input impedance than that of sample B when the matching thickness varies. This feature implies that sample A may have much better microwave absorption performance that a wide microwave absorption band could be obtained.

3.5. Microwave absorption of the core-shell composite particles

The frequency ($f$) dependence of reflection loss (RL) at a certain absorber thickness ($d$) was calculated from complex permeability ($\varepsilon_r = \varepsilon' - j\varepsilon''$) and permittivity ($\mu_r = \mu' - j\mu''$), according to the following formulas [25, 26]:

$$Z_{in} = \sqrt{\frac{\mu_r}{\varepsilon_r}} \tanh \frac{2\pi f d \sqrt{\mu_r \varepsilon_r}}{c} \quad (7)$$

$$RL(dB) = 20\log\left|\frac{Z_{in}-1}{Z_{in}+1}\right| \quad (8)$$

Where $c$ is the velocity of electromagnetic waves in free apace, $Z_{in}$ is the normalized input impedance of absorber. It shows that the reflection loss depends sensitively on the thickness and frequency. According to equation (8), the RL value of -20 dB is comparable to 99% microwave absorption. Generally, the RL value less than -20 dB is considered as adequate microwave absorption.

The *RL* curves as a function of frequency are shown in Fig. 9. All the peaks in the transmission loss spectra shift to the lower frequency with increasing thickness. As illustrated in Fig. 9(a), The RL values are less than -20 dB in the range of 7.9-12.3 GHz over absorber thicknesses of 2.3-3.4 mm and the minimum is -41 dB at the frequency of 9.3 GHz. The frequency bandwidth whose reflection loss is over -20 dB is 4.4 GHz. Therefore, it is reasonably deduced that there exists a critical thickness, at which $Z_{in}$ nearest the point of $α=1$ and $β=0$ in the smith chart, the reflection loss will have the minimum but a finite value. As illustrated in Fig. 9(b), the minimum of RL is -56 dB at the frequency of 12.3 GHz with a critical thickness of 2.47 mm. Meanwhile, the frequency bandwidth whose reflection loss is over -20 dB is narrow. These features illuminate that the microwave absorption properties of these composites are mainly due to the proper impedance match and that is a key factor for microwave absorptions.

## 4. Conclusions

Both of the Ni-SiO$_2$ core-shell structures could have good microwave absorption performance because permittivity/permeability is crucially determined by the percent and the intrinsic EM parameters of the component. Smith chart reveals that the minimum RL value was obtained at a critical thickness where the input EM impedance was reduced and nearest the air impedance. So it could be concluded that the enhanced wide-band microwave absorption of the composite coatings was due to the proper electromagnetic impedance matches, not dependent on the constructed different structures.

## Acknowledgements

This work is supported by the National Natural Science Foundation of China with grant number 11074101, the National Science Fund for Distinguished Young Scholars (50925103). We also thanks for the support from Fundamental Research Funds for the Central Universities (lzujbky-2011-54).

**Figure captions:**

**Fig. 1.** Schematic illustration of Ni coated $SiO_2$ and $SiO_2$ coated Ni composite with core-shell structures.

**Fig. 2.** TEM micrographs of (a) prepared silica spheres, (b) pre-treated silica powders.

**Fig. 3.** SEM images of Ni coated $SiO_2$ core-shell composite at different magnifications (a, b) , Ni particles (c) and $SiO_2$ coated Ni particles with core-shell structures (d).

**Fig. 4.** The XRD patterns of Ni-$SiO_2$ composite powders with reversed core-shell structures.

**Fig. 5.** The magnetic hysteresis loops of Sample A and Sample B.

**Fig. 6.** Frequency dependence of the real part and imaginary part of complex permeability of Sample A and Sample B.

**Fig. 7.** Effective permittivity and permeability is presented as a function of the core component percent in the core-shell structures for samples A and B.

**Fig. 8.** Smith chart for reduced input impedance of the core-shell structured composite for sample A and Sample B.

**Fig. 9.** Microwave reflection loss spectra of the composite powders for Sample A and Sample B at various thicknesses.

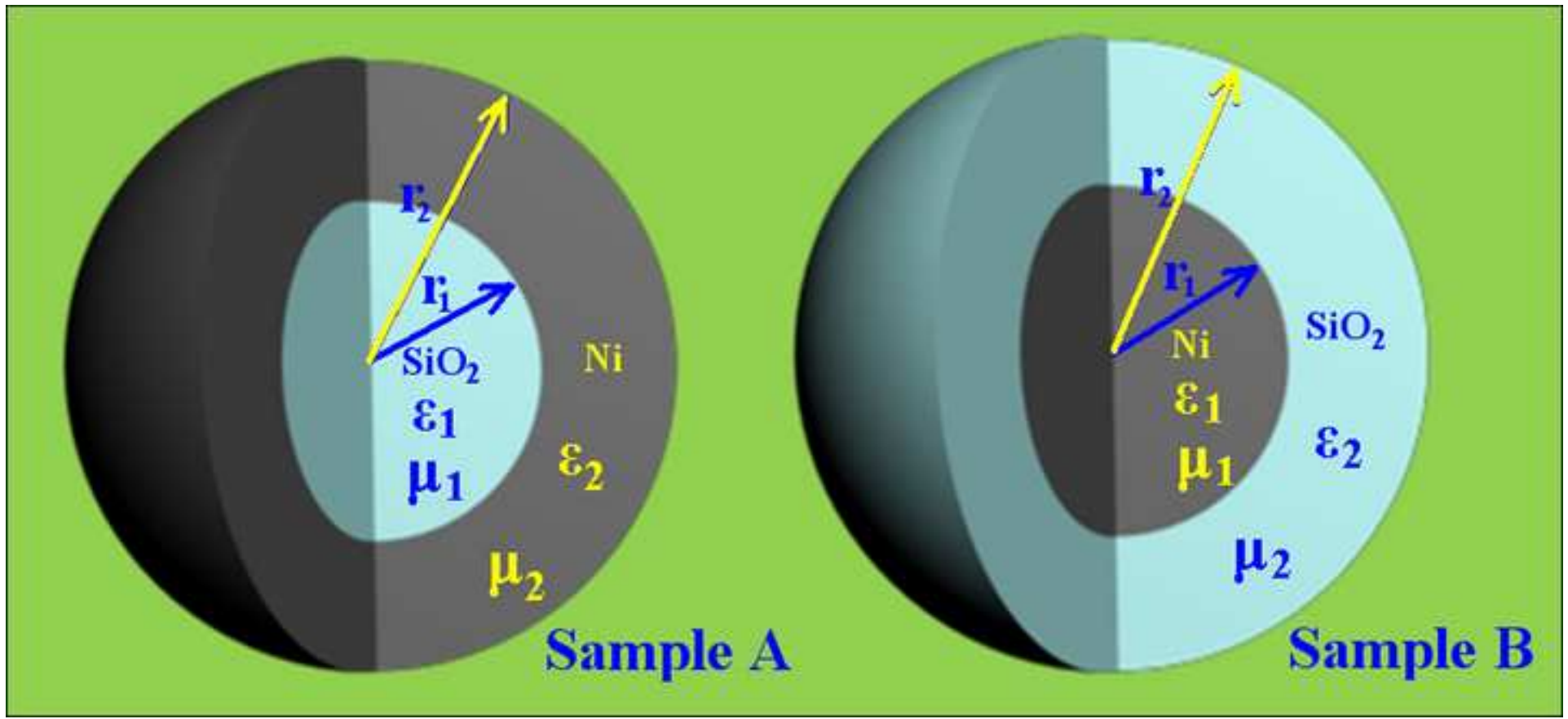

**Figure(2)**
**Click here to download high resolution image**

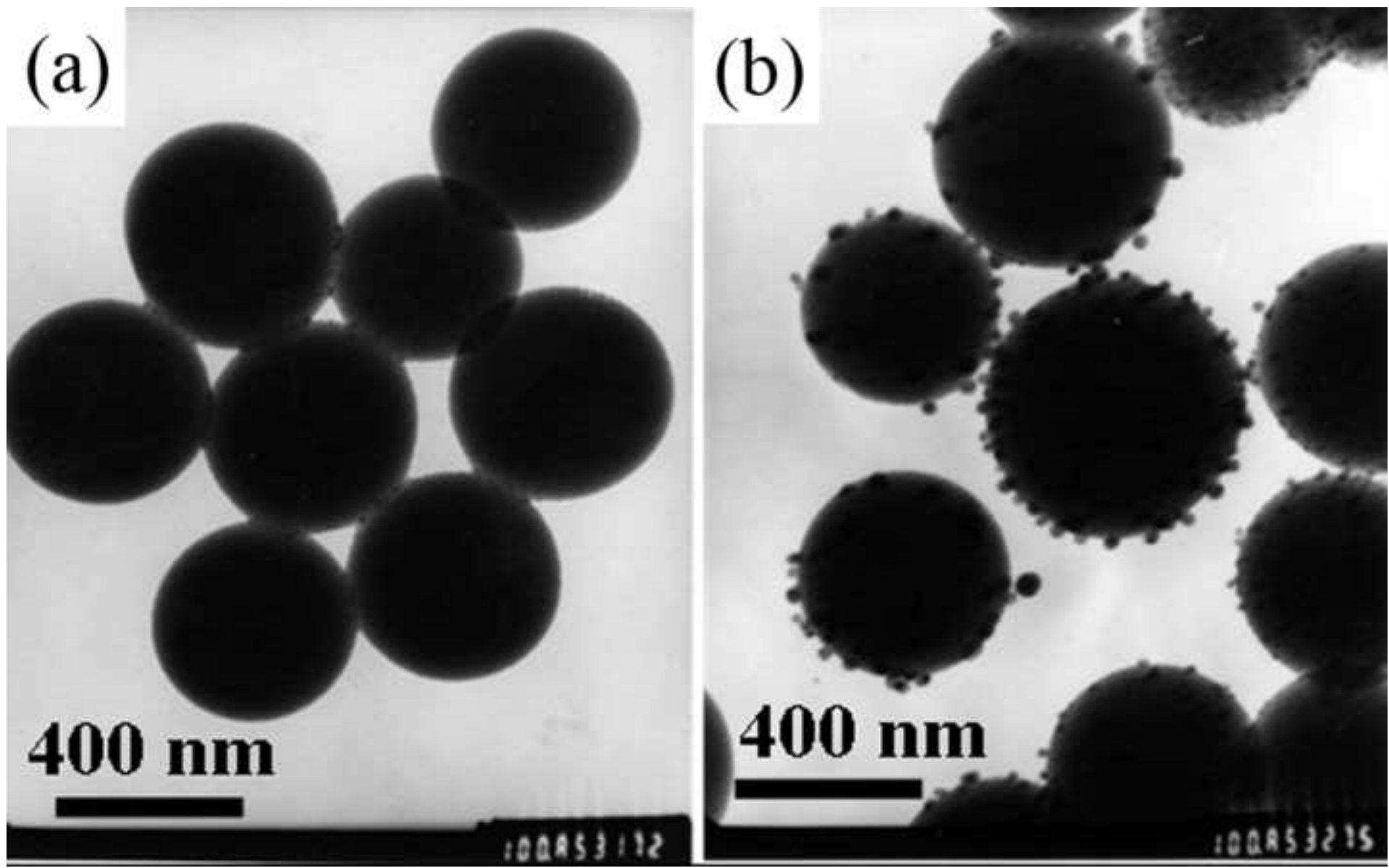

**Figure(3)**
Click here to download high resolution image

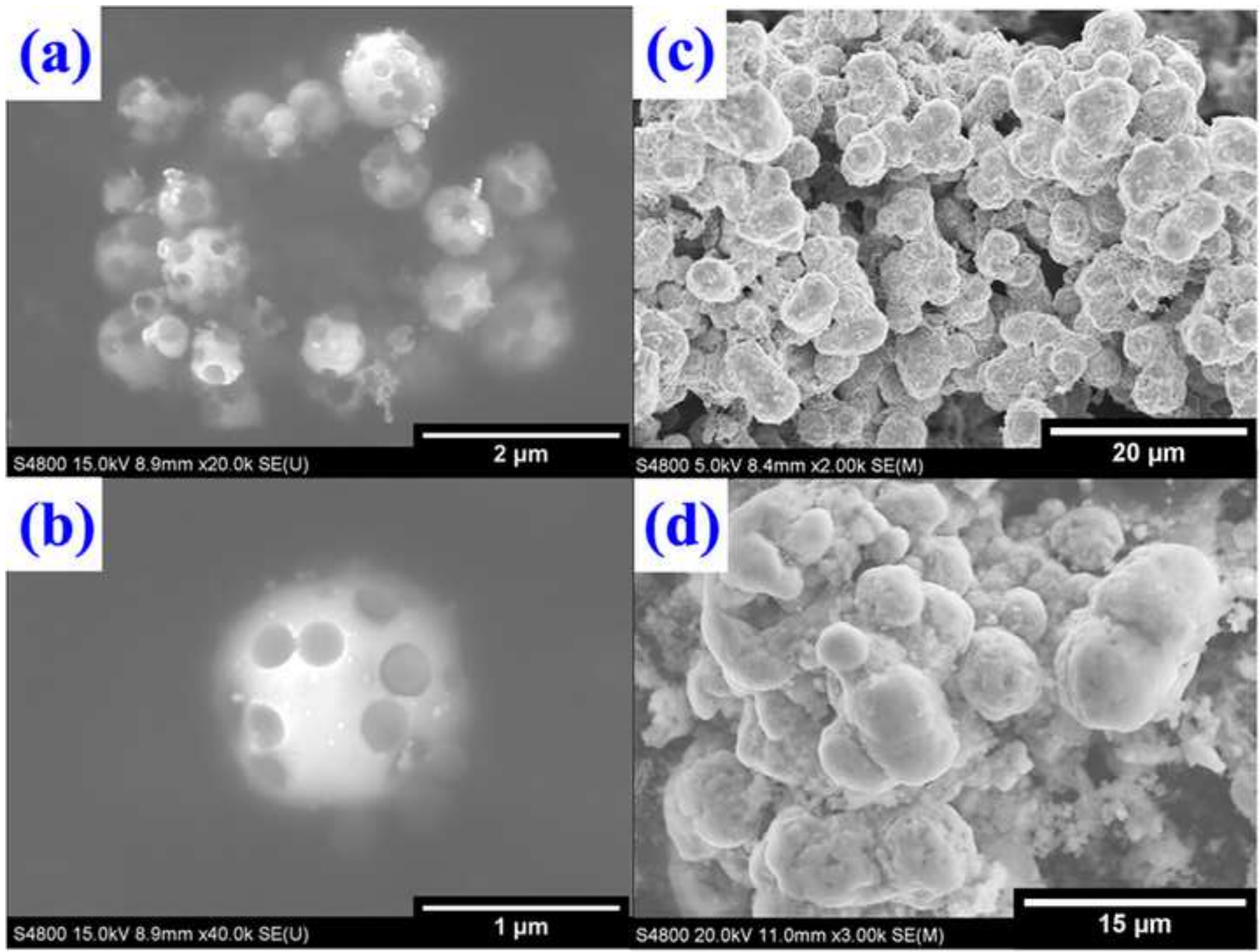

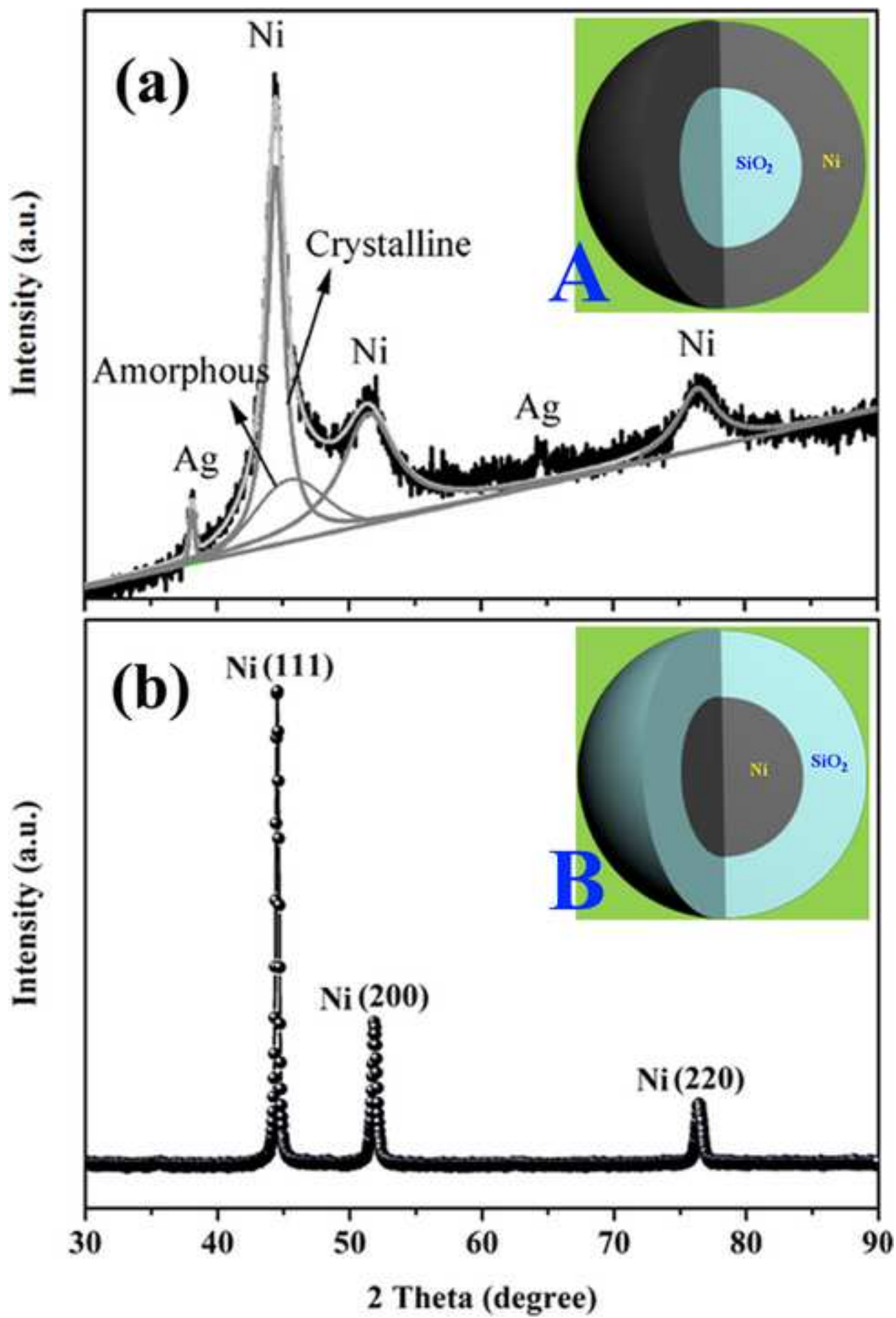

**Figure(5)**
**Click here to download high resolution image**

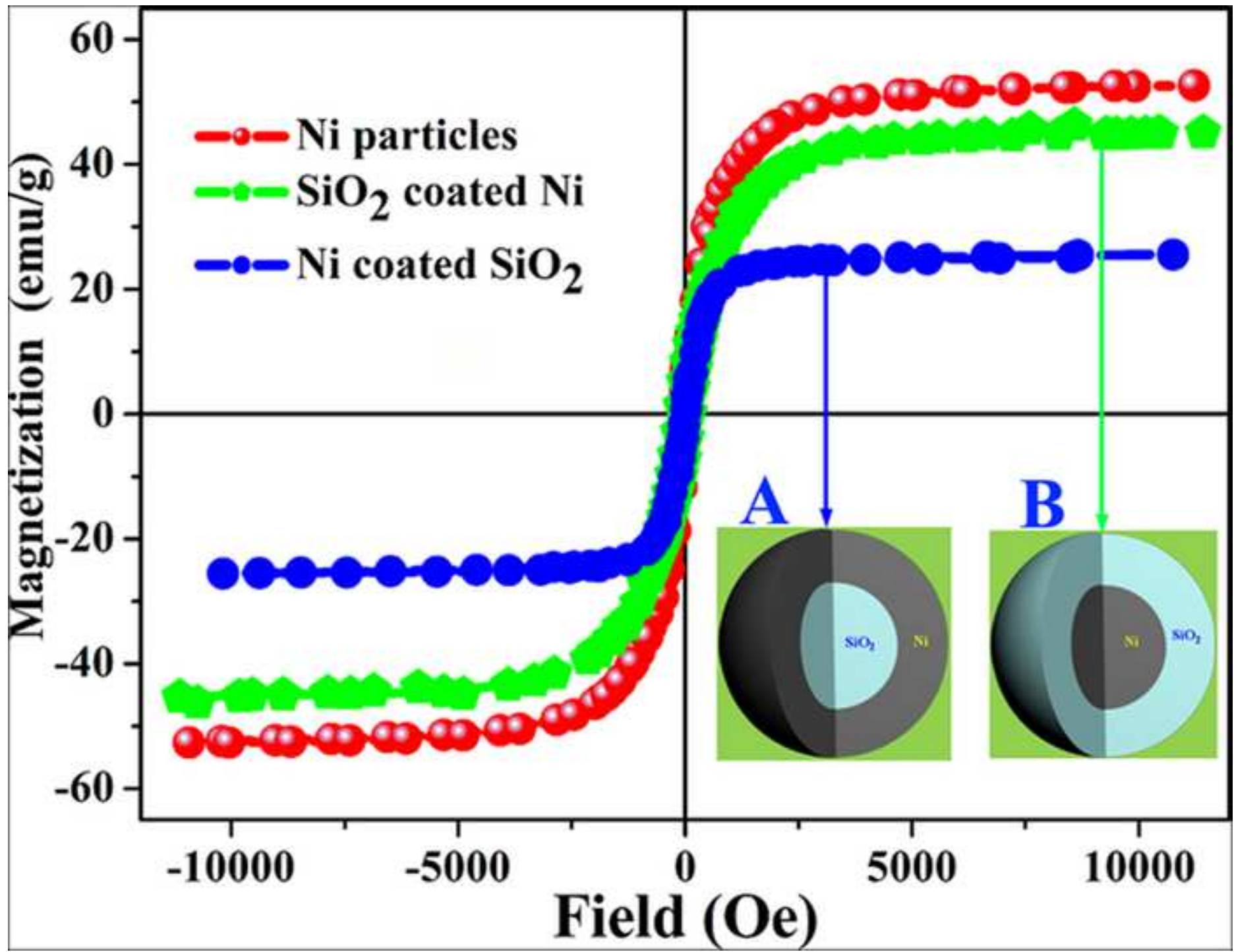

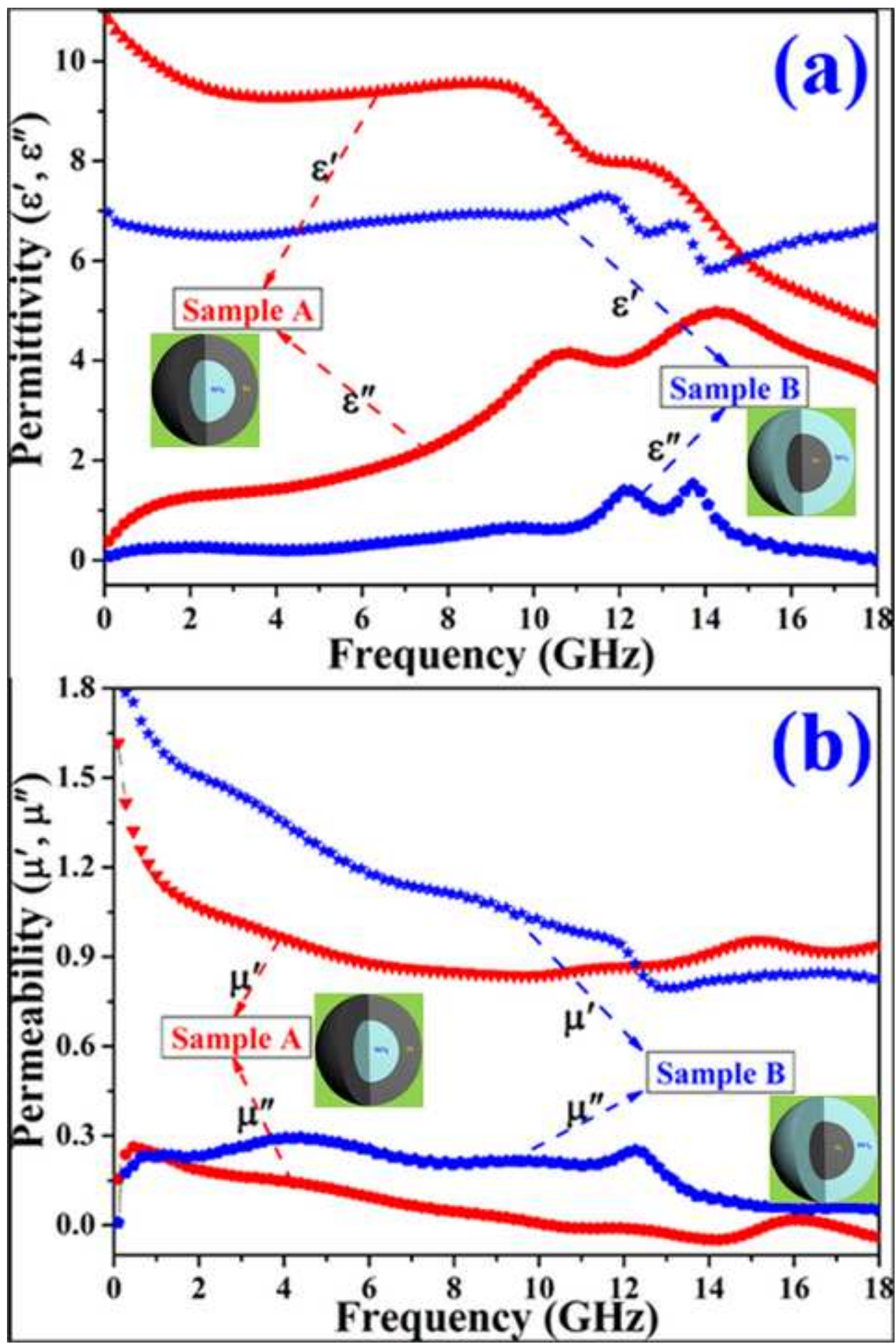

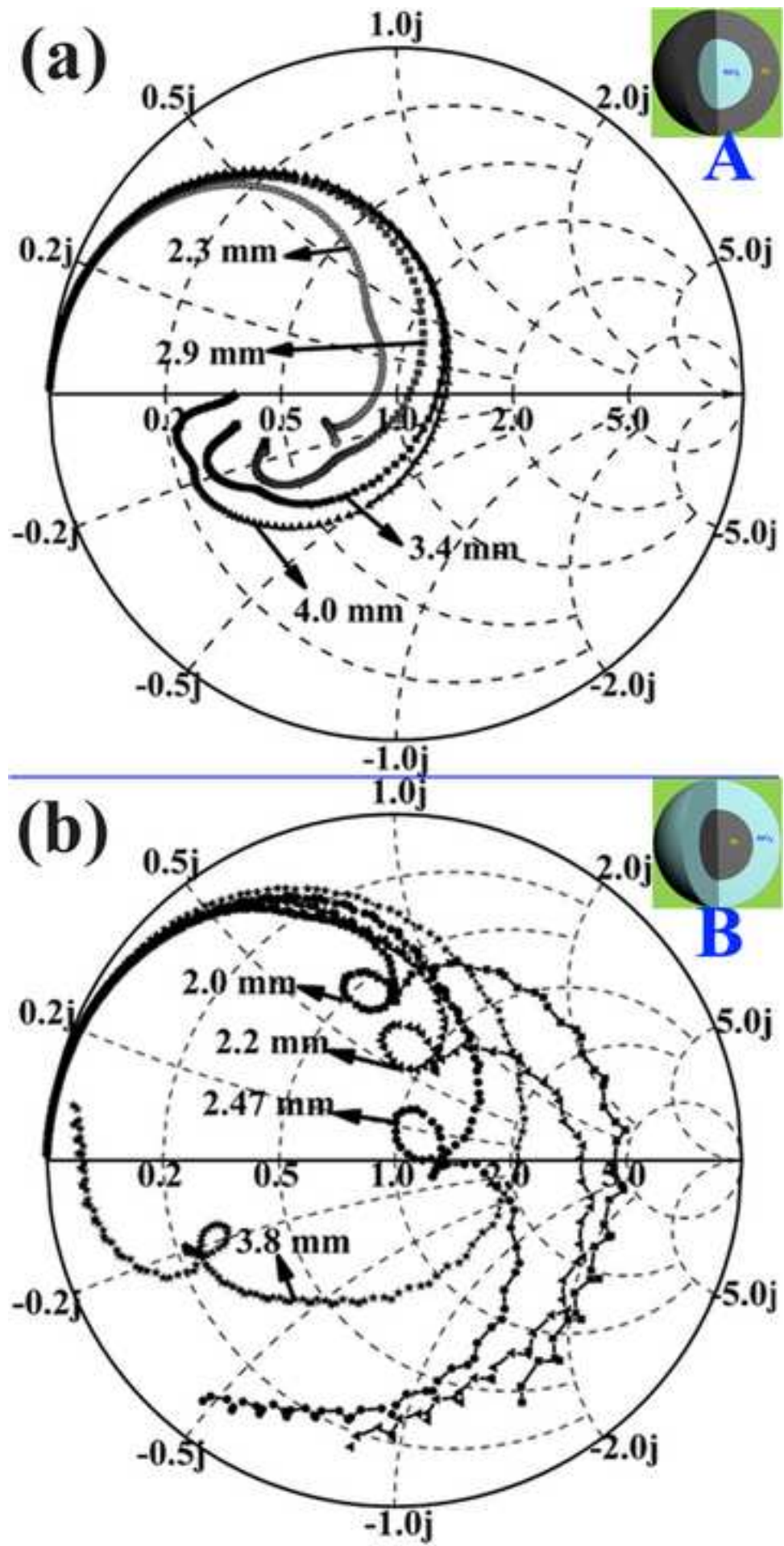

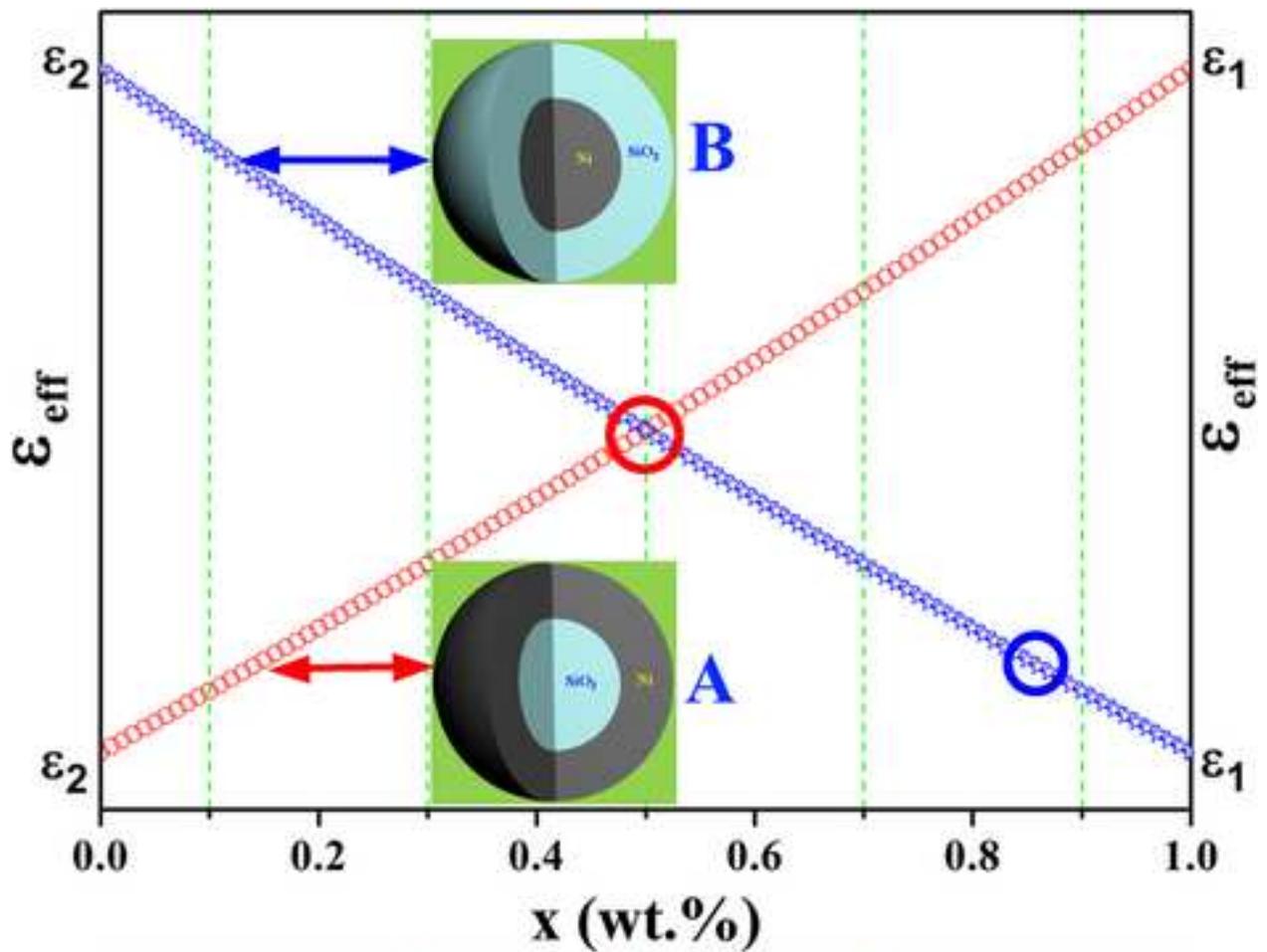
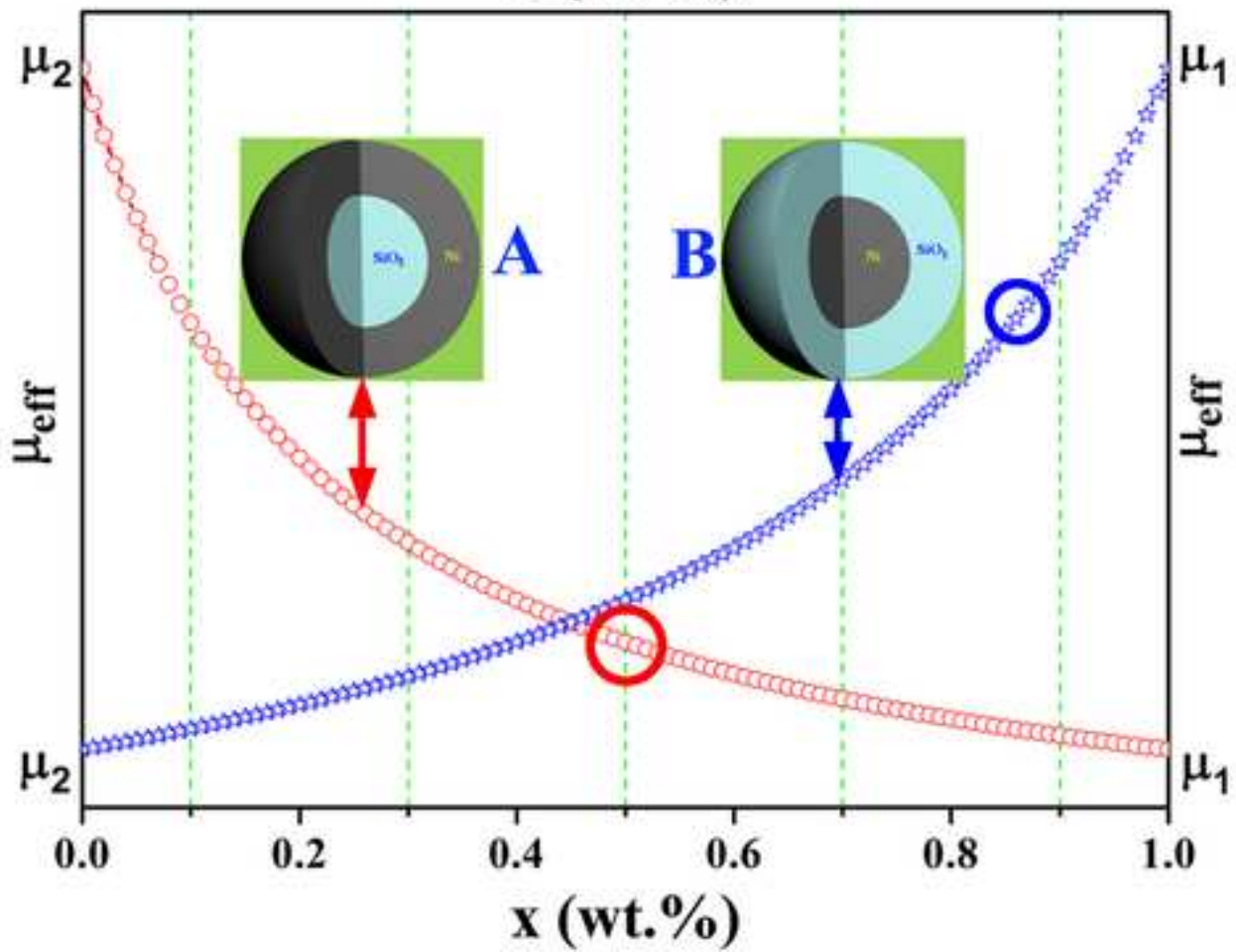

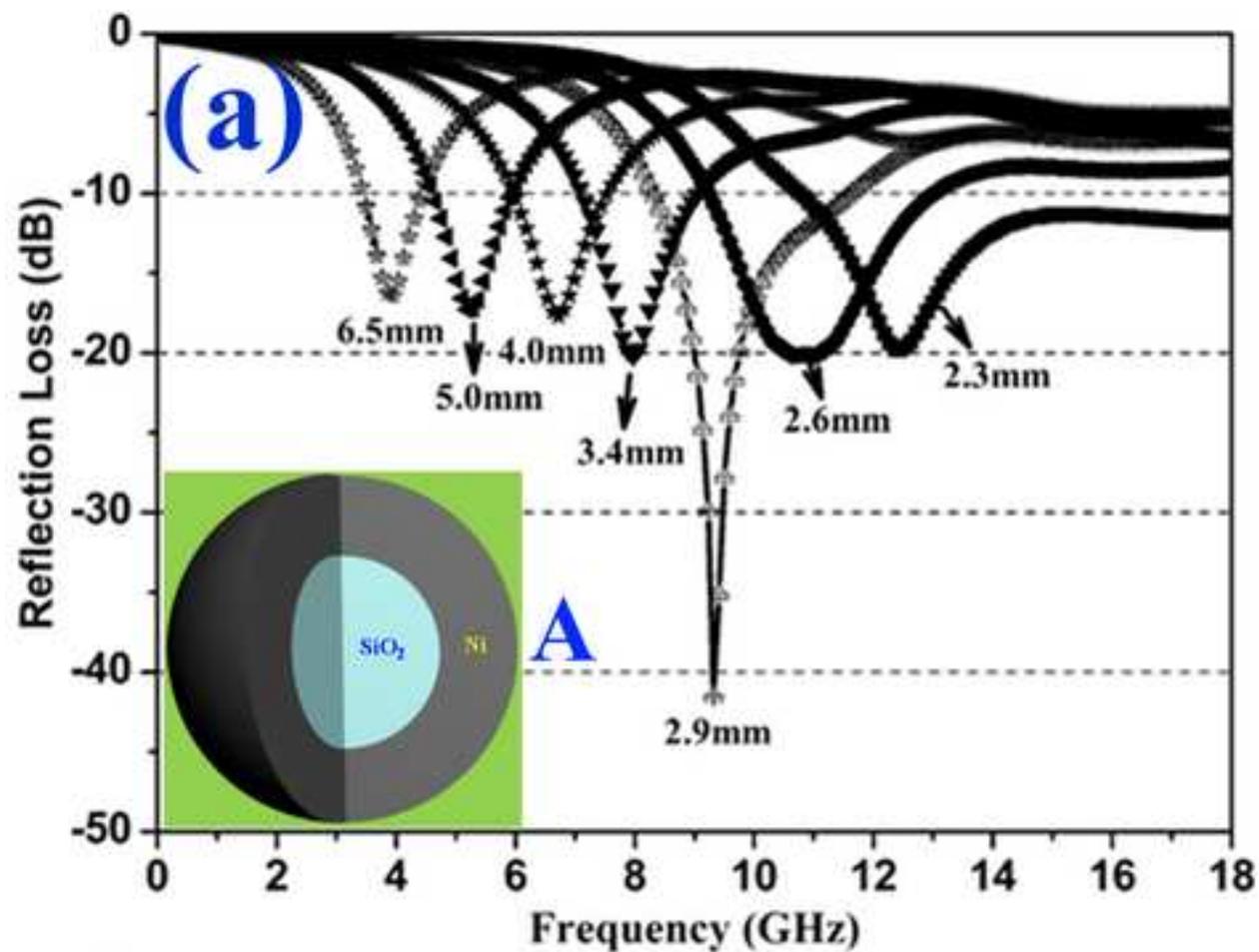
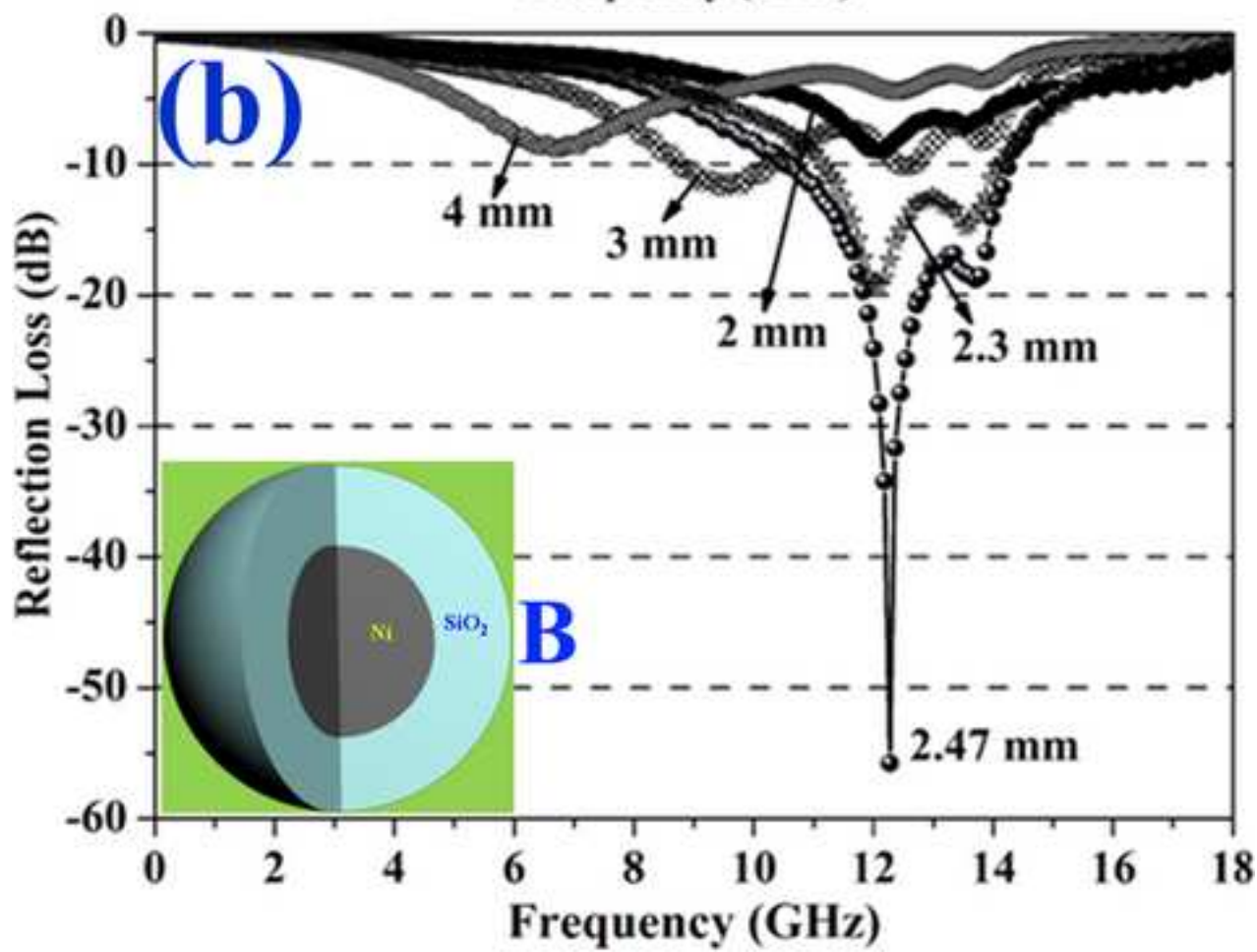